%% file: main.tex
\documentclass[cameraready]{Interspeech}
\usepackage[utf8]{inputenc}
\usepackage[T1]{fontenc}
\usepackage{amsfonts}
\usepackage{microtype}

\title{ZipL-Dialog: Memory-Efficient Long-Form Spoken Dialog Synthesis\\via Latent Flow Matching}

\author[affiliation={1,2}]{Jihwan}{Kim}
\author[affiliation={1}]{Nam Soo}{Kim}

\address{
    $^1$ Department of Electrical and Computer Engineering and INMC, \\ Seoul National University, Seoul, South Korea \\
    $^2$ KT Corporation, Seoul, South Korea
}

\email{jhkim17@hi.snu.ac.kr, nkim@snu.ac.kr}


\begin{document}
\setlength{\abovedisplayskip}{3pt}
\setlength{\belowdisplayskip}{3pt}

\maketitle

\begin{abstract}
Zero-shot dialog TTS benefits from flow-matching, but minute-scale generation on dense mel-spectrograms causes severe memory bottlenecks, often forcing unnatural chunked synthesis. We propose ZipL-Dialog, which shifts conditional flow-matching into a 4x time-compressed (25 Hz) latent space. To preserve acoustic fidelity under compression, we employ a deterministic mel autoencoder with auxiliary mel-domain supervision and optimize the ZipFormer's hierarchical downsampling schedule. Experiments show that ZipL-Dialog reduces maximum peak GPU memory by 11.22x and accelerates inference by 2.23x over the baseline, substantially lowering the memory footprint of single-pass multi-minute dialog synthesis while maintaining perceptual naturalness. Audio samples are available at \url{https://speechdemos.github.io/}.
\end{abstract}
\section{Introduction}
\label{sec:intro}

Recent advances in neural speech synthesis have extended text-to-speech (TTS) beyond short, single-utterance synthesis toward \emph{long-form} and \emph{multi-turn} conversational audio, including podcasts, role-play dialogs, and interactive agents.
In these settings, models must preserve naturalness, speaker and turn consistency, and contextual coherence over horizons approaching minutes, while remaining efficient enough for practical deployment.

A prominent approach to long-form conversational synthesis is \emph{autoregressive} (AR) generation, in which acoustic tokens or frames are produced sequentially (e.g.,~\cite{vibevoice,fireredtts2,soulxpodcast}).
AR models often provide strong quality and robust conditioning, but their inference latency typically increases linearly with output duration because each generation step depends on previously generated outputs.
As a result, synthesizing multi-minute dialog can incur substantial wall-clock latency, which limits responsiveness in interactive applications and scalability in large-scale deployment.

To alleviate AR latency, \emph{non-autoregressive} (NAR) diffusion and flow-based models generate acoustic sequences in parallel through a fixed number of refinement steps or ODE integration.
Recent dialog-capable systems such as CoVoMix, CoVoMix2, and ZipVoice-Dialog (e.g.,~\cite{covomix,covomix2,zipvoice_dialog}) follow this direction and substantially improve throughput over AR generation.
However, when such models operate directly on mel-spectrogram, long-form dialog introduces a different bottleneck: memory grows rapidly with sequence duration.
Under conditional flow matching (CFM)~\cite{cfm}, the model processes the full acoustic sequence at once, so longer audio increases activation storage, intermediate states, and, for transformer-based backbones, attention cost during training and inference.
Consequently, minute-scale dialog often requires aggressive truncation, short-segment training, or chunked generation, which can weaken long-range conversational modeling.

A natural way to reduce this burden is to move generation into a temporally compressed latent space.
Latent diffusion has proven effective in high-resolution image synthesis by learning an autoencoder that compresses the signal before diffusion is applied in the latent domain~\cite{stablediffusion}.
Likewise, flow matching can be performed in latent space to improve efficiency while retaining the flexibility of continuous generative modeling~\cite{latentcfm}.
In speech synthesis, M3-TTS shows that latent acoustic modeling can improve the efficiency of NAR synthesis~\cite{m3tts}.
These results motivate latent-space flow/diffusion for speech; however, extending such methods to multi-minute, multi-turn dialog remains challenging because the latent representation must preserve phonetic and speaker-related detail while supporting much longer conversational context than standard single-utterance TTS.

In this work, we ask: \emph{Can latent-space CFM substantially reduce the memory and runtime cost of long-form dialog synthesis while preserving competitive speech quality?}
To answer this question, we propose \textbf{ZipL-Dialog}, a latent conditional flow-matching framework for efficient zero-shot spoken dialog synthesis.
Instead of modeling frame-level acoustic features directly, ZipL-Dialog performs generation in a time-compressed continuous latent space and decodes the predicted latents back to the acoustic domain.
This design shortens the sequence processed by the flow model and is particularly beneficial for minute-scale dialog, where memory becomes a dominant constraint.

A key challenge is that aggressive temporal compression can remove phonetic and speaker-relevant detail.
We address this with two design choices.
First, we adopt a deterministic mel autoencoder with auxiliary mel-domain supervision, which better preserves acoustic detail under strong compression in our setting than a variational latent formulation.
Second, because standard hierarchical downsampling is not necessarily optimal once the input is already temporally compressed, we redesign the ZipFormer downsampling schedule to better balance contextual aggregation and local resolution.

The contributions of this work are threefold:
\begin{itemize}
    \item We propose \textbf{ZipL-Dialog}, a latent CFM framework for long-form spoken dialog synthesis that reduces sequence-length-dependent memory and runtime by operating in a 25\,Hz time-compressed latent space.
    \item We show that a \textbf{deterministic latent representation} combined with an \textbf{auxiliary mel-domain reconstruction loss} preserves acoustic detail and intelligibility more effectively than a variational alternative under strong temporal compression in our setting.
    \item We analyze hierarchical downsampling for compressed latent sequences and identify a \textbf{ZipFormer downsampling schedule} that provides a better efficiency--quality trade-off for long-form dialog synthesis.
\end{itemize}

\section{Related Work}
\label{sec:related}

\subsection{Flow Matching for Zero-Shot Dialog TTS}
Recent zero-shot TTS systems have increasingly adopted diffusion and flow-matching frameworks to enable non-autoregressive speech synthesis with strong naturalness and conditioning fidelity.
For conversational speech, models such as CoVoMix and CoVoMix2 \cite{covomix, covomix2} have shown that conditional flow matching (CFM) can model multi-turn structure, speaker alternation, and diverse dialog prosody within a unified generation framework.
ZipVoice-Dialog \cite{zipvoice_dialog} further demonstrates that a ZipFormer-based CFM backbone can produce highly natural zero-shot dialog speech while maintaining robust contextual modeling across turns.
Among these prior systems, ZipVoice-Dialog is most closely related to our work.
It performs dialog generation directly in a dense frame-level acoustic domain using a Vocos-compatible mel-spectrogram \cite{vocos}.
While this design provides strong synthesis quality, operating on full-length frame sequences remains increasingly expensive as dialog duration grows.
Under minute-scale generation, activation storage, intermediate states, and sequence-processing cost all increase substantially with output length, making memory and runtime practical bottlenecks even in non-autoregressive settings.

\subsection{Latent Acoustic Generation and Compression--Fidelity Trade-offs}
Compressing speech into a lower-rate latent representation is a promising strategy for reducing the sequence-length burden of generative modeling.
In speech synthesis, M3-TTS \cite{m3tts} combines a mel-latent codec with a diffusion-transformer backbone and shows that latent acoustic modeling can improve the efficiency of non-autoregressive synthesis.

However, maintaining speech quality under strong temporal compression remains challenging.
Many latent generative systems rely on variational autoencoders (VAEs), whose probabilistic regularization can over-smooth local spectral and phonetic detail when the temporal resolution is aggressively reduced, especially under limited model capacity.
This issue becomes more critical in long-form dialog, where the representation must preserve not only intelligibility but also speaker-related cues and turn-level continuity over much longer horizons than standard single-utterance TTS.

Prior latent acoustic generation studies have largely focused on single-utterance or general speech synthesis settings, leaving the efficiency--quality trade-off for multi-minute, multi-turn dialog relatively underexplored.
These observations motivate exploring latent representations that remain temporally compact while better preserving fine acoustic detail, as well as reconsidering architectural choices once the input sequence has already been compressed.

\section{Method}
\label{sec:method}

ZipL-Dialog performs dialog generation in a temporally compressed latent acoustic space rather than directly in the frame-level mel domain.
Given a frame-level mel-spectrogram, we first encode it into a lower-rate continuous latent sequence, apply masked conditional flow matching in the latent domain, and finally decode the predicted latents back to the mel domain for waveform synthesis.

\subsection{Deterministic Mel Autoencoder}
\label{subsec:autoencoder}

To reduce the sequence length handled by the flow model, we learn an acoustic autoencoder that maps a frame-level mel-spectrogram to a time-compressed latent sequence.
Instead of adopting a variational latent formulation, we use a deterministic bottleneck.
In our setting, strong temporal compression combined with variational regularization tended to over-smooth local phonetic detail, whereas a deterministic latent representation allowed the model capacity to focus more directly on reconstruction fidelity.

Let $\mathbf{Y}\in\mathbb{R}^{T\times F}$ denote the input mel-spectrogram, where $T$ is the number of time frames and $F$ is the number of mel bins.
The encoder $\mathcal{E}$ first applies a 2D convolutional patch embedding with kernel and stride $(F, r)$, spanning the full frequency axis and $r$ consecutive frames, to produce a sequence of temporal tokens.
After a depthwise 1D convolution for local positional mixing, the token sequence is processed by a stack of Transformer encoder layers and projected through a 1D convolutional bottleneck to a latent sequence $\mathbf{Z}\in\mathbb{R}^{T'\times D}$, where $T'=T/r$.

The decoder $\mathcal{D}$ maps $\mathbf{Z}$ back to the hidden dimension, refines the features using ConvNeXt-style residual blocks, and restores the original frame rate through transposed-convolution upsampling layers.
In all experiments, we use a temporal compression factor of $r=4$ and a latent dimension of $D=100$, which yields a 25\,Hz latent sequence when the frame-level acoustic representation is extracted at 100\,Hz.

\subsection{Masked Latent Conditional Flow Matching}
\label{subsec:latentcfm}

\begin{figure}[t]
  \centering
  \includegraphics[width=\linewidth]{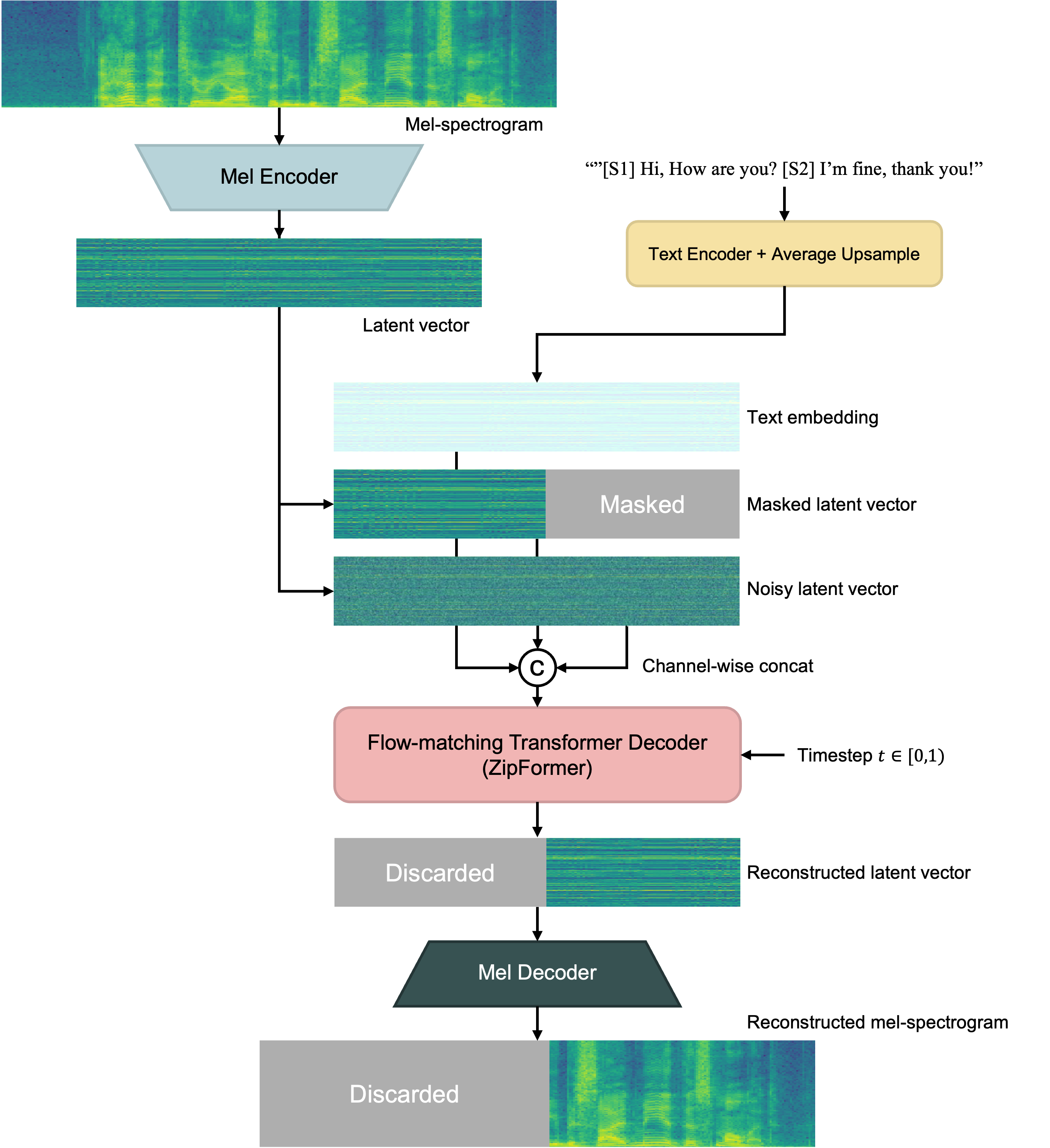}
  \caption{Training overview of ZipL-Dialog. A ground-truth mel-spectrogram is encoded into a $4\times$ time-compressed latent sequence (25\,Hz). The ZipFormer decoder predicts the CFM velocity on the masked target region, conditioned on text embeddings and noisy latents. The model is jointly optimized with a latent velocity loss and an auxiliary mel-domain reconstruction loss.}
  \label{fig:training}
\end{figure}

Fig.~\ref{fig:training} illustrates the training procedure of ZipL-Dialog.
Given a ground-truth mel-spectrogram $\mathbf{Y}$, we first obtain its compressed latent sequence
$\mathbf{Z}=\mathcal{E}(\mathbf{Y})\in\mathbb{R}^{T'\times D}$ using the acoustic autoencoder.
We then divide $\mathbf{Z}$ into an observed prefix context region and a target region to be generated.
This partition is represented by a binary mask $\mathbf{m}\in\{0,1\}^{T'}$ (broadcast over channels), where $\mathbf{m}_t=1$ denotes the target region.

To construct the conditioning inputs, we first form a masked latent sequence
\begin{equation}
\mathbf{Z}^{\mathrm{mask}}=(1-\mathbf{m})\odot\mathbf{Z}+\mathbf{m}\odot\mathbf{z}_{\mathrm{mask}},
\end{equation}
where $\mathbf{z}_{\mathrm{mask}}$ is the mask embedding used for the target region.
Following conditional flow matching, we sample $\boldsymbol{\epsilon}\sim\mathcal{N}(\mathbf{0},\mathbf{I})$ and $t\sim\mathcal{U}(0,1)$, and define the noisy latent state as
\begin{equation}
\mathbf{Z}_t=(1-\mathbf{m})\odot\mathbf{Z}+\mathbf{m}\odot\left((1-t)\boldsymbol{\epsilon}+t\mathbf{Z}\right).
\end{equation}
Thus, the context region remains clean, while only the target region follows the linear interpolation path between Gaussian noise and the clean latent.

The input text, augmented with explicit speaker and turn tokens (e.g., \texttt{[S1]} and \texttt{[S2]}), is encoded by a text encoder.
Its token-level representations are then average-upsampled to latent-frame-level embeddings
$\mathbf{E}\in\mathbb{R}^{T'\times d}$ aligned to the latent sequence length.
The ZipFormer-based flow decoder takes the concatenation of $\mathbf{Z}_t$, $\mathbf{Z}^{\mathrm{mask}}$, and $\mathbf{E}$ as input and predicts the target velocity $\hat{\mathbf{v}}_{\theta}$.

We optimize the model with a masked velocity-matching objective on the target region:
\begin{equation}
\mathcal{L}_{\mathrm{vel}}
=
\left\lVert
\mathbf{m}\odot
\left(
\hat{\mathbf{v}}_{\theta}-(\mathbf{Z}-\boldsymbol{\epsilon})
\right)
\right\rVert_2^2 .
\end{equation}

To further preserve acoustic detail, we recover a denoised latent estimate
\begin{equation}
\hat{\mathbf{Z}}
=
(1-\mathbf{m})\odot\mathbf{Z}
+
\mathbf{m}\odot
\left(
\mathbf{Z}_t+(1-t)\hat{\mathbf{v}}_{\theta}
\right),
\end{equation}
decode it back to the mel domain as $\hat{\mathbf{Y}}=\mathcal{D}(\hat{\mathbf{Z}})$, and apply an auxiliary masked reconstruction term
\begin{equation}
\mathcal{L}_{\mathrm{mel}}
=
\left\lVert
\mathbf{m}^{\uparrow}\odot
\left(
\hat{\mathbf{Y}}-\mathbf{Y}
\right)
\right\rVert_2^2 ,
\end{equation}
where $\mathbf{m}^{\uparrow}$ denotes the mask upsampled to the mel frame rate.
The final training objective is
\begin{equation}
\mathcal{L}
=
\mathcal{L}_{\mathrm{vel}}
+
\lambda \mathcal{L}_{\mathrm{mel}}.
\end{equation}
The observed prefix context is used only as conditioning and is excluded from the training loss.

At inference time, the observed dialog prefix is encoded into the latent domain, the masked target region is initialized from Gaussian noise, and the learned velocity field is integrated from $t=0$ to $t=1$ to generate the target latent sequence.
The generated latent sequence is then decoded to a mel-spectrogram and converted to waveform using the neural vocoder.

\subsection{ZipFormer Downsampling Design}
\label{subsec:downsampling}

The default ZipFormer hierarchy used in ZipVoice-Dialog was designed for dense frame-level acoustic inputs.
When the same downsampling schedule is applied after $4\times$ temporal compression, each higher-level token covers a substantially longer time span, which can degrade short-phoneme resolution and harm local acoustic fidelity.
Therefore, the default hierarchy is not directly optimal in the latent setting.

Our goal is to balance three factors: (i) preserving sufficient local acoustic resolution, (ii) expanding the receptive field for stable text--speech alignment and long-range dialog modeling, and (iii) retaining the efficiency benefits of hierarchical processing.
To this end, we compare representative downsampling schedules spanning no, moderate, and aggressive temporal reduction, and select the one that provides the best efficiency--quality trade-off.
As shown in Section~\ref{subsec:ablation}, the default schedule is suboptimal for compressed latent sequences, and a milder hierarchy yields better overall behavior in our setting.

\section{Experiments}
\label{sec:experiments}

We conduct all experiments on English data, matching the language coverage of the large-scale training mixture and the evaluation protocols of prior dialog TTS benchmarks.

\subsection{Training Datasets}
\label{subsec:data}

\noindent\textbf{Backbone pretraining.}
Following the ZipVoice training recipe, we pretrain the backbone on a mixture of large-scale English speech corpora: HiFiTTS2 (\textasciitilde31.7k hours) \cite{hifitts2}, Emilia-English (\textasciitilde20k hours) \cite{emilia}, and LibriTTS (\textasciitilde585 hours) \cite{libritts}.

\noindent\textbf{Dialog fine-tuning.}
Following ZipVoice-Dialog, we fine-tune on the English subset (\textasciitilde5k hours) of the 6.8k-hour OpenDialog training corpus, which is constructed from in-the-wild spoken-dialog data.

\subsection{Implementation Details}
\label{subsec:impl}

\noindent\textbf{Shared configuration.}
All models are trained on four NVIDIA A100 (80GB) GPUs. For fair comparison, ZipL-Dialog and the ZipVoice-Dialog baseline use identical Vocos-compatible 100-bin mel-spectrograms (24\,kHz, 100\,Hz frame rate). Unless otherwise noted, the optimizer and learning-rate schedules follow the official ZipVoice-Dialog implementation.

\noindent\textbf{Mel autoencoder.}
The deterministic autoencoder ($\sim$34\,M parameters) is trained on the shared English speech mixture using 3\,s random audio crops. It is optimized via AdamW (learning rate $10^{-4}$ with warmup) and a batch size of 200 for 200k steps. Its weights are then frozen during ZipL-Dialog training.

\noindent\textbf{ZipL-Dialog training.}
The backbone ($\sim$123\,M parameters) incorporates our optimized \texttt{[1,1,2,1,1]} downsampling schedule. Training utilizes dynamic batching (up to 1{,}200\,s per batch) in two stages. First, the backbone is pretrained for 200k steps, filtering data to utterances under 30\,s for memory control and stability. Subsequently, it is fine-tuned on the dialog data for 100k steps. The auxiliary mel-domain reconstruction loss weight is set to $\lambda=0.5$.

\subsection{Evaluation Datasets}
\label{subsec:eval_data}

We evaluate ZipL-Dialog on two dialog-oriented benchmarks: (i) the OpenDialog test set used in ZipVoice-Dialog, and (ii) the CoVoMix2 dialog test set proposed in \cite{covomix2}.
The CoVoMix2 dialog test set contains 1k dialog transcripts sampled from DailyDialog \cite{dailydialog} and uses acoustic prompts from LibriSpeech test-clean \cite{librispeech}.

\subsection{Baseline Models}
\label{subsec:baselines}

We compare ZipL-Dialog against two strong baselines from complementary generative paradigms:
\begin{itemize}[leftmargin=*,nosep]
    \item \textbf{ZipVoice-Dialog (NAR baseline) \cite{zipvoice_dialog}:} the most closely related frame-level non-autoregressive dialog TTS baseline built on the same architectural family.
    \item \textbf{VibeVoice 1.5B (AR baseline) \cite{vibevoice}:} a large-scale autoregressive speech synthesis model included to contrast the efficiency--quality trade-off at multi-minute dialog duration.
\end{itemize}

\subsection{Evaluation Metrics}
\label{subsec:metrics}

\textbf{Efficiency metrics.}
All runtime and memory measurements are obtained end-to-end with a batch size of 1 on a single NVIDIA A100 40GB GPU, encompassing prompt processing, latent generation, mel decoding, and waveform synthesis. We report inference time, average RTF, average peak GPU memory, and the maximum peak memory observed across the test set.

\noindent\textbf{Quality metrics.}
Following ZipVoice-Dialog \cite{zipvoice_dialog}, WER is computed using WhisperD \cite{parakeet} after standard text normalization and speaker-tag removal. For cpSIM, we utilize Pyannote diarization \cite{pyannote} and WavLM-ECAPA embeddings \cite{wavlm,ecapa} to find the maximum average cosine similarity under possible speaker permutations. We also report perceptual quality via UTMOS \cite{utmos}.

\noindent\textbf{Inference settings.}
Both ZipL-Dialog and the ZipVoice-Dialog baseline employ a 16-step Euler ODE solver with classifier-free guidance. For VibeVoice, we evaluate using its official open-source inference implementation.
\subsection{Main Results}

\input{efficiency}
\textbf{Efficiency and Scalability.}
Table~\ref{tab:efficiency} shows that ZipL-Dialog is substantially more efficient than both baselines. It is 43--47$\times$ faster than VibeVoice and reduces maximum peak GPU memory by up to 11.22$\times$ relative to ZipVoice-Dialog, indicating that 25\,Hz latent-space flow matching markedly lowers the runtime and memory cost of long-form dialog synthesis.

\input{quality}
\noindent\textbf{Quality and Efficiency--Quality Trade-off.}
Table~\ref{tab:quality} shows that ZipL-Dialog does not match the uncompressed baseline on all objective metrics: ZipVoice-Dialog yields lower WER on both sets, and ZipVoice-Dialog/VibeVoice obtain higher cpSIM on CoVoMix2/OpenDialog. Still, ZipL-Dialog achieves the best or tied-best UTMOS on both benchmarks, suggesting that overall naturalness is preserved despite modest losses in local phonetic and speaker-similarity detail.

\subsection{Ablation Study: Latent Types, Auxiliary Loss, and Downsampling}
\label{subsec:ablation}

For a single-speaker zero-shot TTS evaluation, we follow the F5-TTS protocol \cite{f5tts} and report WER, SIM-o, and UTMOS on the LibriSpeech-PC test-clean set \cite{librispeechpc}. WER is computed using a HuBERT-large ASR model \cite{hubert}. Note that these absolute scores are not directly comparable to our dialog benchmarks due to differences in the evaluation setup.

\input{ablation}

\noindent\textbf{Results and Analysis.}
Table~\ref{tab:ablation} details our ablation results. Although the VAE latent slightly favors speaker similarity (SIM-o 0.518 vs. 0.489), our deterministic AE substantially improves intelligibility (WER 3.634\%) and naturalness (UTMOS 3.982), making it the optimal choice to prevent phonetic blurring under 4$\times$ compression. Furthermore, incorporating the auxiliary loss ($\mathcal{L}_{\mathrm{mel}}$) consistently improves all metrics over the baseline AE. Finally, our moderate downsampling schedule (\texttt{[1,1,2,1,1]}) drastically outperforms both the no-downsampling and the overly aggressive default schedules, proving that compressed latents still require carefully balanced hierarchical processing.

\section{Conclusion}
\label{sec:conclusion}
We proposed ZipL-Dialog, an efficient latent conditional flow-matching framework for long-form spoken dialog synthesis. By moving generation to a 25\,Hz deterministic latent space and redesigning the ZipFormer downsampling schedule, ZipL-Dialog substantially reduces the memory and runtime cost of minute-scale synthesis. On the CoVoMix2 and OpenDialog test sets, it achieves up to 11.22$\times$ lower maximum peak GPU memory and up to 2.23$\times$ faster inference than ZipVoice-Dialog, while maintaining competitive perceptual quality. Although temporal compression introduces modest trade-offs in objective intelligibility and speaker-similarity metrics, ZipL-Dialog attains the best or tied-best UTMOS on both benchmarks, demonstrating a strong efficiency--quality trade-off for long-form dialog generation. Future work will investigate improved latent modeling and reconstruction objectives to better recover local phonetic detail without sacrificing efficiency.

\section{Generative AI Use Disclosure}
The authors utilized OpenAI ChatGPT and Google Gemini to improve the English phrasing, readability, and grammar of this manuscript. All content and ideas remain the sole original work of the authors.

\bibliographystyle{IEEEtran}
\bibliography{mybib}

\end{document}

%% file: efficiency.tex
\begin{table}[t]
\centering
\caption{Efficiency results measured on a single NVIDIA A100 40GB GPU (lower is better).}
\label{tab:efficiency}
\footnotesize
\setlength{\tabcolsep}{3pt}
\begin{tabular}{lccc}
\toprule
\textbf{Metric} & \textbf{Ours} & \textbf{ZipVoice-Dialog} & \textbf{VibeVoice 1.5B} \\
\midrule
\multicolumn{4}{c}{\textbf{CoVoMix2 test set} (avg=32.878\,s, max=178.891\,s)} \\
\midrule
Inference time (s) & \textbf{1.075} & 2.396 & 50.160 \\
RTF & \textbf{0.056} & 0.089 & 1.455 \\
Peak memory (GB) & \textbf{1.10} & 4.01 & 5.49 \\
Max peak memory (GB) & \textbf{3.23} & 36.21 & 6.20 \\
\midrule
\multicolumn{4}{c}{\textbf{OpenDialog test set} (avg=26.029\,s, max=65.141\,s)} \\
\midrule
Inference time (s) & \textbf{1.045} & 1.477 & 45.069 \\
RTF & \textbf{0.052} & 0.069 & 1.558 \\
Peak memory (GB) & \textbf{0.97} & 2.03 & 5.34 \\
Max peak memory (GB) & \textbf{1.30} & 5.94 & 5.41 \\
\bottomrule
\end{tabular}
\end{table}

%% file: quality.tex
\begin{table}[t]
\centering
\caption{Quality results (WER lower is better; cpSIM/UTMOS higher is better).}
\label{tab:quality}
\footnotesize
\setlength{\tabcolsep}{3pt}
\begin{tabular}{lccc}
\toprule
\textbf{Metric} & \textbf{Ours} & \textbf{ZipVoice-Dialog} & \textbf{VibeVoice 1.5B} \\
\midrule
\multicolumn{4}{c}{\textbf{CoVoMix2 test set}} \\
\midrule
WER (\%) & 5.203 & \textbf{4.229} & 4.959 \\
cpSIM & 0.444 & \textbf{0.493} & 0.485 \\
UTMOS & \textbf{3.523} & 3.477 & \textbf{3.523} \\
\midrule
\multicolumn{4}{c}{\textbf{OpenDialog test set}} \\
\midrule
WER (\%) & 5.362 & \textbf{3.550} & 12.979 \\
cpSIM & 0.304 & 0.346 & \textbf{0.350} \\
UTMOS & \textbf{3.198} & 3.089 & 2.312 \\
\bottomrule
\end{tabular}
\end{table}

%% file: ablation.tex
\begin{table}[t]
\centering
\caption{Controlled ablation in a single-speaker zero-shot TTS setting trained on LibriTTS and evaluated on LibriSpeech-PC. SIM-o denotes speaker similarity in the single-speaker setting; its absolute values are not directly comparable to the multi-speaker cpSIM scores in Tables~\ref{tab:efficiency} and \ref{tab:quality}.}
\label{tab:ablation}
\footnotesize
\setlength{\tabcolsep}{2pt}
\begin{tabular}{lccc}
\toprule
\textbf{Setting} & \textbf{WER$\downarrow$} & \textbf{SIM-o$\uparrow$} & \textbf{UTMOS$\uparrow$} \\
\midrule
\multicolumn{4}{l}{\textit{(1) Latent Representation}} \\
VAE & 6.535 & \textbf{0.518} & 3.877 \\
\textbf{AE (Ours)} & \textbf{3.634} & 0.489 & \textbf{3.982} \\
\midrule
\multicolumn{4}{l}{\textit{(2) Auxiliary Loss (with AE)}} \\
w/o $\mathcal{L}_{mel}$ & 4.024 & 0.485 & 3.896 \\
\textbf{w/ $\mathcal{L}_{mel}$ (Ours)} & \textbf{3.634} & \textbf{0.489} & \textbf{3.982} \\
\midrule
\multicolumn{4}{l}{\textit{(3) Downsampling (with AE, w/ $\mathcal{L}_{mel}$)}} \\
A (\texttt{[1,1,1,1,1]}) & 51.964 & 0.418 & 3.676 \\
\textbf{B (\texttt{[1,1,2,1,1]}, Ours)} & \textbf{3.634} & \textbf{0.489} & \textbf{3.982} \\
C (\texttt{[1,2,4,2,1]}, Def.) & 27.432 & 0.357 & 3.671 \\
\bottomrule
\end{tabular}
\end{table}